\documentclass[showpacs,twocolumn,amsmath,amssymb,amsfonts,floatfix,aps,prb, 10pt]{revtex4-1}

\usepackage{graphicx, color,rotating} 
\usepackage{dcolumn} 
\usepackage{bm} 
\usepackage{epsfig}
\usepackage{bbm}
\usepackage{dsfont}
\usepackage{relsize}
\usepackage{braket}
\usepackage{tikz}
\usepackage{soul}
\usepackage{stmaryrd}
\usepackage[plainpages=false, pdfpagelabels, bookmarksopen, pdfborder={0 0 0}, linkcolor=blue, linkbordercolor={0 0 1}]{hyperref}
\usepackage{feynmp}

\begin{document}

\title{Time local master equation connecting the Born and Markov approximations}

\author{Christian Karlewski}
\email{christian.karlewski@kit.edu}
\author{Michael Marthaler}
\affiliation{Institut f\"ur Theoretische Festk\"orperphysik and DFG-Center for Functional Nanostructures, Karlsruhe Institute of Technology, D-76131 Karlsruhe, Germany}



\begin{abstract}
We present an exact expansion of the master equation for an open quantum system.
The resulting equation is time local and enables us to calculate clearly defined higher order corrections to the Born-Markov approximation. 
In particular, we show that non-Markovian terms are of the same order of magnitude as higher order terms in the system-bath coupling.
As a result we emphasize that analyzing non-Markovian behaviour of a system implies going beyond Born approximation.
Additionally, we address with this approach the initial state problem occurring in non-Markovian master equations.
\end{abstract}

\maketitle

\section{Introduction}
The quantum master equation (QME) for open quantum systems is of huge interest for many different fields of research in physics \cite{SWB-107920557}.
It describes a quantum system with a small number of degrees of freedom which is coupled to another system with many degrees of freedom commonly called 'the bath'. 
The difficult task to characterize a quantum mechanical system which is connected with its environment is e.g. crucial for our understanding of
the transition from quantum to  classical physics \cite{1-s2.0-0003491683902026-mai,406043a0}. 
Because of the generality of the formalism the field of applications is large. It is used for atoms interacting with electromagnetic fields \cite{416206a}, non-equilibrium quantum mechanics \cite{PhysRevA.73.042108,PhysRevA.87.062117}, chemical reactions \cite{RevModPhys.62.251}, tunneling processes \cite{1-s2.0-0003491683902026-mai}, circuit quantum electrodynamics \cite{0309049,PhysRevA.69.062320,PhysRevLett.107.093901} and many other systems.

The QME is an equation of motion of the reduced density matrix of the system of interest $\rho$
\begin{align}
  \dot{\rho}(t)=-i[H_S(t),\rho]+\int_{t_0}^t \text{d}t' \Sigma(t-t')\rho(t'). \label{eQME}
\end{align}
The effect of the bath is contained in the kernel $\Sigma(t-t')$ and the coherent evolution of the system is given by the first commutator. Because this equation is time non-local and the kernel can be arbitrary complicated, a general solution is often rather difficult or not practicable. So, many different approaches have been developed to solve this equation and we want to present a short overview. An exact treatment of the problem was made by a path integral solution \cite{1.468244,Weiss2012}. This solution comes by the cost that it is resource consuming and only small systems can be treated. Another exact treatment of the QME was made by Shi and Geva\cite{JCPSA6-119-23-12063-1} using a extension of the Nakajima-Zwanzig projection method (NZPM). The NZPM introduces two superoperators which projects density matrices on the system or the bath. Shi and Geva could divide the kernel in three parts which can be solved exactly using e.g. path integral methods, but needs less resources. Both methods can treat arbitrary system-bath couplings, but are rather complicated in use. Most other methods rely on a expansion parameter and some assumptions on the system. We want to mention here the time-convolutionless method \cite{SWB-107920557,1-s2.0-S0003491601961524-main,Mitra2004}, the T-Matrix approach \cite{Koch2005,Elste2007,Lueffe2008,Akera1999,Golovach2004}, the Keldysh-contour approach \cite{PhysRevB.50.18436,PhysRevLett.76.1715,PhysRevLett.78.4482,PhysRevB.68.115105} and the text-book Bloch-Redfield master equation \cite{SWB-107920557,Wangsness1953,Bloch1957,Redfield1965}, which are all compared in the comprehensive work of Timm \cite{PhysRevB.77.195416}.   

The two most common approximations solving the QME are the Markov and Born approximation. The Markov approximation implies that the state at time $t+\partial t$, where $\partial t$ is an infinitesimal time step, only depends on its state at time $t$. This is true for a 'memory-less' bath which means that the bath correlation function decays fast compared to the system dynamics. But of course, this is not legitimate for all open quantum systems \cite{PhysRevLett.109.170402} and the initial state problem is one consequence. If the future state of the system depends on its past, it is not possible to start the simulation with an out of equilibrium initial state at time $t_0$. The moment the system is initialized, the actual initial state correlations which depend on the way the density matrix was brought to its current state have been neglected which is 
only valid for a Markovian system.

If system and bath are coupled weakly, the Born approximation can be used. It is the lowest order expansion in the coupling between the bath and the quantum mechanical system of interest. A correct expansion of the QME is thus an expansion in correlation time and coupling strength. Accordingly, using the famous combination of the two approximations, the Born-Markov approximation, can be a keen confinement and as mentioned before much effort has been made to go beyond these approximations.

In the last 15 years the investigation of non-Markovian behaviour has become more and more important because it appears that for many quantum mechanical system of interest the Markov-approximation is not valid \cite{JCPSA6-119-23-12063-1,Weiss2012}. But to simplify calculations and often to get analytic results, the Born-approximation was still used \cite{PhysRevB.71.035318,PhysRevLett.93.130406,JChemPhys_111_3365,PhysRevLett.111.180602,PhysRevE.88.052127,PhysRevLett.104.177403}. Our method focuses on the validity of these approaches.

We developed an expansion of the QME in terms which can be addressed as orders beyond the Born and Markov approximation. The expansion gives us the possibility to exactly establish a connection between Born and Markov approximation and to show their dependencies. All terms are expanded in the coupling between bath and system, divided by the correlation time of the bath. 
This means that it is possible to estimate the order of magnitude of the next term of each of the approximations. As our main result we find that using the Born approximation by investigating non-Markovian behaviour of quantum systems is in general not reasonable. This method is most useful if the explicit integral kernels do not have to be calculated, but a order of magnitude estimation is needed. As an example of this procedure, we show that a specific term can be assigned to the initial state correlations and the error of the initial state problem can be quantified.

This paper is structured as follows: in the section \ref{cNonMark} we develop the Markovian expansion and in section \ref{cDia} an expansion in the coupling strength to the bath in diagrammatic form. The section \ref{cFull} combines these two expansions. We present then as an example the investigation of the spin-boson model in section \ref{cSpin} and a more general analysis of the initial state problem in section \ref{cIni}. In the end we conclude in section \ref{cCon}.

\section{Non-Markovianity}
\label{cNonMark}
Starting from the QME (\ref{eQME}) at an initial time $t_0\rightarrow-\infty$, our first step is to get a time local equation of the reduced density matrix. This is possible with an expansion in terms of derivatives of the reduced density matrix and primitive integrals of the kernel. We will show later that there is a small parameter which makes this a meaningful expansion. A related approach was used by Rojek {\it et al.} in the case of pumping quantum dots \cite{pssb201350213}. The first step is to transform eq.(\ref{eQME}) to the interaction picture ($\hbar=1$), i.e. 
\begin{align}
  A_I(t)=&U^\dagger(t_0,t)AU(t_0,t),\\
  U(t_0,t)=&\mathcal{T}\,e^{i\int_{t_0}^t \text{d}t'H(t')},
\end{align}
where $\mathcal{T}$ is the time-ordering operator.
The resulting QME is given by
\begin{align}
  \dot{\rho}_I(t)=\int_{t_0}^t \text{d}t' \Sigma_I(t-t')\rho_I(t'). \label{eQMEI}  
\end{align}
The expansion is achieved by integration by parts of eq.(\ref{eQMEI}) where an upper index will mark the primitive integral of a function. The primitive integral of our kernel is given by
\begin{align}
 \Sigma_I^{(k+1)}(t-t')=\int_{\infty}^{t-t'}\text{d}t''\Sigma_I^{(k)}(t'').
\end{align}
With these ingredients eq.(\ref{eQMEI}) becomes 
\begin{align}
  \dot{\rho}_I=&-\left[ \int_{\infty}^{t-t'}\text{d}t''\Sigma_I^{(0)}(t'')\rho_I(t') \right]_{t'=-\infty}^t\notag\\
    &+\int_{-\infty}^t \text{d}t'\Sigma_I^{(1)}(t-t')\dot{\rho}_I(t').
\end{align}
The term in squared brackets evaluated at minus infinity vanishes and the term with $t'=t$ can be identified as the Markov approximation. 
Integration by parts gives the next terms in the Markov expansion. An efficient way of writing this can be achieved by introducing $\mathcal{S}^{(k)}=\left(\int_{0}^{\infty}\text{d}t'\Sigma_I^{(k)}(t') \right)$ and $\rho_{I(k)}(t)$ as the $k^{\text{th}}$ derivative of $\rho_I(t)$. The dynamics of this $k^{\text{th}}$ derivative is given by
\begin{align}
  k>0;\quad\rho_{I(k)}(t)=\sum_{l=k-1}^{\infty}\mathcal{S}^{(1-k+l)}\rho_{I(l)}(t).
  \label{eDerME}
\end{align}
We are interested in eq.(\ref{eQMEI}) which is the equation for the first derivative 
\begin{align}
  \rho_{I(1)}(t)=\sum_{l=0}^{\infty}\mathcal{S}^{(l)}\rho_{I(l)}(t).
  \label{eExpME}
\end{align}
This equation is exact if all summands are taken into account meaning no approximation is used and this equation equals eq. (\ref{eQMEI}). Now, we recursively insert eq.(\ref{eDerME}) in eq.(\ref{eExpME}). The idea is to reduce the $k^{\text{th}}$ derivative of $\rho_{I(k)}$ till it reaches $\rho_{I(0)}$ (to keep the equation compact we write $\rho_{I(k)}$ without its argument)
\begin{align}
  \rho_{I(1)}&=\underbrace{\mathcal{S}^{(0)}}_{A_1}\rho_{I(0)}+\sum_{l=1}^{\infty}\mathcal{S}^{(l)}\sum_{m=l-1}^{\infty}\mathcal{S}^{(1-l+m)}\rho_{I(m)},\\
  \rho_{I(1)}&=A_1\rho_{I(0)}+\underbrace{\mathcal{S}^{(1)}\mathcal{S}^{(0)}}_{A_2}\rho_{I(0)}\notag\\
    &+\sum_{l=1}^{\infty}\mathcal{S}^{(l)}\sum_{m=l}^{\infty}\mathcal{S}^{(1-l+m)}\rho_{I(m)}+\sum_{l=2}\mathcal{S}^{(l)}\mathcal{S}^{(0)}\rho_{I(l-1)}.
\end{align}
Inserting eq.(\ref{eDerME}) to all orders, eq.(\ref{eQMEI}) can be written as 
\begin{align}
  \rho_{I(1)}(t)&=\left(\sum_{n=1}^{\infty}A_n  \right)\rho_{I(0)}(t).
  \label{eTIME}
\end{align}
The composition of $A_n$ made up out of different $\mathcal{S}^{(k)}$ is a combinatorial problem
\begin{align}
  A_n=\sum_{m}\left[\prod_1^n \mathcal{S}^{(f_m(n))}\right].
\end{align}
The rules for $f_m(n)$ to get the correct terms are:
\begin{itemize}
  \item The sum of the indices $f_m(n)$ for all $\mathcal{S}^{(f_m(n))}$ of one term must be $n-1$, $f_m(n)\in\mathbb{N}_0$.
  \item Assign each $\mathcal{S}^{(f_m(n))}$ from right to left a position index p. The sum of the indices $f_m(n)$ from 1 to a given position p must be smaller p.
  \item All terms that fulfill the two rules above must be summed ({\scriptsize $\sum_m$}).
\end{itemize}
As an illustrating example we show the calculation of $A_3$
\begin{align}
  \left.\begin{matrix}
  \mathcal{S}^{(2)}\mathcal{S}^{(0)}\mathcal{S}^{(0)}&\Rightarrow{\color{green} \checkmark}\\
  \mathcal{S}^{(0)}\mathcal{S}^{\color{red} (2)}\mathcal{S}^{(0)}&\Rightarrow{\color{red} \lightning}\\
  \mathcal{S}^{(0)}\mathcal{S}^{(0)}\mathcal{S}^{\color{red}(2)}&\Rightarrow{\color{red} \lightning}\\
  \mathcal{S}^{(1)}\mathcal{S}^{(1)}\mathcal{S}^{(0)}&\Rightarrow{\color{green} \checkmark}\\
  \mathcal{S}^{(1)}\mathcal{S}^{(0)}\mathcal{S}^{\color{red} (1)}&\Rightarrow{\color{red} \lightning}\\
  \mathcal{S}^{(0)}\mathcal{S}^{(1)}\mathcal{S}^{\color{red} (1)}&\Rightarrow{\color{red} \lightning}
  \end{matrix}\right\}A_3=\mathcal{S}^{(2)}\mathcal{S}^{(0)}\mathcal{S}^{(0)}+\mathcal{S}^{(1)}\mathcal{S}^{(1)}\mathcal{S}^{(0)}.
\end{align}

The equation (\ref{eTIME}) is an exact expansion of the QME (\ref{eQMEI}) and we name this the Markov expansion. It implies that all derivatives of $\rho_{I(0)}$ equal zero in the steady state solution and the Markovian and non-Markovian solution are identical. This can easily be seen by the condition $\Sigma^{I(0)}\rho_{I(0)}=0$ for the steady state solution.
To find the small parameter in eq.(\ref{eTIME}) we have to investigate the time dependence of our kernel and the coupling to the bath.

\section{Diagrammatic expansion}
\label{cDia}
So far we haven't specified the form of our kernel. In principle the kernel contains the whole information about the interaction with the bath. However, in practice we calculate the kernel using an expansion in the system bath coupling. We call this the Born expansion. It is convenient to visualize this expansion by using the diagrammatic expansion on the Keldysh-contour \cite{PhysRevB.50.18436, PhysRevB.74.085305}. If all orders in the coupling are taken into account the QME for the reduced density matrix $\rho_I$ of the system is given by eq.(\ref{eQMEI}). The only approximation made for this equation is that the full density matrix of system and bath $\rho^I_{SB}$ at an initial time $t_0$ can be written as a direct product of the reduced density matrix of the system $\rho_I=\text{tr}_B\left\{\rho^I_{SB}\right\}$ and the bath $\rho^I_{B}=\text{tr}_S\left\{\rho^I_{SB}\right\}$
\begin{align}
  \rho^I_{SB}(t_0)=\rho_I(t_0)\otimes\rho^I_{B}(t_0),
\end{align}
which is in principle valid for the limit $t_0\rightarrow-\infty$ but is not the case for a finite $t_0$ (e.g. $t_0=0$) as shown later.

To investigate the kernel of equation (\ref{eQMEI}), we have to specify how our system looks like. We consider a very general model of a quantum system coupled to a bath. The Hamiltonian we want to investigate is separable in three parts
\begin{align}
  H=H_S+H_B+H_C.
\end{align}
The system Hamiltonian $H_S$ defines the system of interest, the part $H_B$ is the bath Hamiltonian and $H_C$ is the coupling between them. We write the eigenstates and eigenvalues of $H_S$ as $H_S\ket{q}=E_q\ket{q}$. The coupling is of the form
\begin{align}
  H_C=g_c\cdot\sum_i s_iX_i,
\end{align}
with the coupling strength $g_c$, an operator from the system Hilbert space $s_i$  and from the bath Hilbert space $X_i$. The coupling enters explicitly in our kernel $\Sigma_I(t-t')$.
Here, $\Sigma_I(t-t')$ is the self-energy given by all the possible irreducible diagrams on the Keldysh contour
\begin{align}
  \Sigma_I=&\underbrace{
    \begin{tikzpicture}[anchor=base,baseline=8pt]
    \coordinate (A) at (0,0);
    \coordinate (B) at (1,0);
    \coordinate (C) at (0,0.7);
    \coordinate (D) at (1,0.7);
    \draw[line width=1.0pt] (A) -- (B);
    \draw[line width=1.0pt] (A) -- (D);
    \draw[line width=1.0pt] (D) -- (C);
    \fill (A) circle (2pt);
    \fill (D) circle (2pt);
    \end{tikzpicture}+
  \begin{tikzpicture}[anchor=base,baseline=8pt]
  \coordinate (A) at (0,0);
  \coordinate (B) at (1,0);
  \coordinate (C) at (0,0.7);
  \coordinate (D) at (1,0.7);
  \draw[line width=1pt] (A) -- (B);
  \draw[line width=1pt] (B) -- (C);
  \draw[line width=1pt] (D) -- (C);
  \fill (B) circle (2pt);
  \fill (C) circle (2pt);
  \end{tikzpicture}+
  \begin{tikzpicture}[anchor=base,baseline=8pt]
  \coordinate (A) at (0,0);
  \coordinate (B) at (1,0);
  \coordinate (C) at (0,0.7);
  \coordinate (D) at (1,0.7);
  \draw[line width=1pt] (A) -- (B);
  \draw[line width=1pt] (A) to [bend left=60](B);
  \draw[line width=1pt] (D) -- (C);
  \fill (A) circle (2pt);
  \fill (B) circle (2pt);
  \end{tikzpicture}+
  \begin{tikzpicture}[anchor=base,baseline=8pt]
  \coordinate (A) at (0,0);
  \coordinate (B) at (1,0);
  \coordinate (C) at (0,0.7);
  \coordinate (D) at (1,0.7);
  \draw[line width=1pt] (A) -- (B);
  \draw[line width=1pt] (C) to [bend right=60](D);
  \draw[line width=1pt] (D) -- (C);
  \fill (C) circle (2pt);
  \fill (D) circle (2pt);
  \end{tikzpicture}}_{\Sigma_{1}}\notag\\
  &+\underbrace{
  \begin{tikzpicture}[anchor=base,baseline=8pt]
    \coordinate (A) at (0,0);
    \coordinate (B) at (1,0);
    \coordinate (C) at (0,0.7);
    \coordinate (D) at (1,0.7);
    \coordinate (E) at (0.6,0.0);
    \coordinate (F) at (0.3,0.7);
    \draw[line width=1.0pt] (A) -- (B);
    \draw[line width=1.0pt] (A) -- (D);
    \draw[line width=1.0pt] (E) -- (F);
    \draw[line width=1.0pt] (D) -- (C);
    \fill (A) circle (2pt);
    \fill (D) circle (2pt);
    \fill (E) circle (2pt);
    \fill (F) circle (2pt);
    \end{tikzpicture}+
 \begin{tikzpicture}[anchor=base,baseline=8pt]
  \coordinate (A) at (0,0);
  \coordinate (B) at (1,0);
  \coordinate (C) at (0,0.7);
  \coordinate (D) at (1,0.7);
  \coordinate (E) at (0.6,0.0);
  \coordinate (F) at (0.3,0.7);
  \draw[line width=1pt] (A) -- (B);
  \draw[line width=1pt] (C) to [bend right=60](D);
  \draw[line width=1.0pt] (E) -- (F);
  \draw[line width=1pt] (D) -- (C);
  \fill (C) circle (2pt);
  \fill (D) circle (2pt);
  \fill (E) circle (2pt);
  \fill (F) circle (2pt);
  \end{tikzpicture}+...}_{\Sigma_{2}}\quad+\cdots
\end{align}
A line on the upper or lower contour is a free time propagation of the density matrix. A contraction containing two vertices is given by 
\begin{align}
  \gamma_{qq'\bar{q}\bar{q}'}^{ij}(t,t')=g_c^2\braket{\bar{q}|s^I_i(t')|q}\braket{q'|s^I_j(t)|\bar{q}'}\langle X^I_i(t')X^I_j(t)\rangle_{B}. \label{eCon}
\end{align}
The time dependence of the last vertex of the diagram is set to the time of the reduced density matrix $t$. In the QME it is necessary to integrate over all the other free vertices over time taking into account the time ordering of the vertices. This gives for $l$ contractions $2l-1$ time integrals.

In the standard Born approximation we only keep the lowest order, i.e. only a single contraction. This expansion can then be written in numbers of contractions $k$ represented by a lower index
\begin{align}
  \Sigma_I(t,t')=\Sigma_{I1}(t,t')+\Sigma_{I2}(t,t')+...+\Sigma_{Ik}(t,t')+...\,.
\end{align}

\section{Full expansion}
\label{cFull}
We now have a closer look at the time dependence of the kernel. The time dependence of the system operators $s^I_i(t)$ can be absorbed in the evaluation of the bath correlation function (see eq.(\ref{eCon})). We want to estimate the bath with one characteristic parameter, the minimum decay rate $\gamma_{min}$ or maximum correlation time and an exponential decay $C^{ij}(t)=\langle X_i(0)X_j(t)\rangle\propto\exp(-\gamma_{min}t)$. In general, our method is also useable for other correlation functions for which the order of magnitude for each integration can be estimated. Our system contains three relevant energy scales: the characteristic energy scale of the small quantum system $\Delta E$, the coupling strength $g_c$ and the correlation rate $\gamma_{min}$. The $\Delta E$ enters the kernel also in exponential form. Therefore, each integration of the kernel yields a factor that is of the order $1/(\gamma_{min}+\Delta E)$. To simplify the terms we approximate this order of magnitude by $\approx1/\gamma_{min}$. This is possible in the limit of small relevant energy scales of the system compared to $\gamma_{min}$. If the relevant system energy $\Delta E$ is larger than the correlation time, the order of magnitude of can be approximated by $1/\Delta E$, but the method itself is still applicable. The Markov approximation is based on the assumption that $\Delta E$ is much smaller than $\gamma_{min}=1/\tau_{min}$, the system decays fast. The Born approximation should be valid if the coupling is weak, i.e. $\Delta E\gg g_c$. We will see that a combination of this two parameters is the expansion parameter in our expansion.

The previously introduced $\mathcal{S}^{(k)}$ get another index for the number of contractions, i.e. $S^{(k)}=\sum_l S_l^{(k)}$ with $\mathcal{S}^{(k)}_{l}=\left(\int_{-\infty}^{t}\text{d}t'\Sigma^{(k)}_{Il}(t-t') \right)$. The order of magnitude of $\mathcal{S}^{(k)}_{l}$ contains two factors. First, the number of contractions $l$ generates a factor $g_c^{2l}$. Second, the number of integrals given by the diagrams $2l-1$ and the primitive integrals $k$. Hence, the small parameter which our expansion is based on is of the order $\mathcal{O}\left(g_c^{2l}/\gamma_{min}^{2l-1+k}\right)$. All together, our final expansion of the QME (\ref{eQMEI}) is 
\begin{align}
   \rho_{I(1)}(t)&=\left(\sum_{n=1}^{\infty}\sum_{m}\left[\prod_1^n \sum_l\mathcal{S}^{(f_m(n))}_l\right]  \right)\rho_{I(0)}(t). \label{eFQME}
\end{align}
We show as an example the expansion up to $\mathcal{O}(g_c^6/\gamma_{min}^5)$
\begin{align}
    \rho_{I(1)}=&\left(\mathcal{S}^{(0)}_1+\mathcal{S}^{(0)}_2+\mathcal{S}^{(0)}_3+\right.\notag\\
    &\phantom{(}\mathcal{S}^{(1)}_1\mathcal{S}^{(0)}_1+\mathcal{S}^{(1)}_2\mathcal{S}^{(0)}_1+\mathcal{S}^{(1)}_1\mathcal{S}^{(0)}_2+ \label{eTIHg6}\\   
    &\left.\mathcal{S}^{(2)}_1\mathcal{S}^{(0)}_1\mathcal{S}^{(0)}_1+\mathcal{S}^{(1)}_1\mathcal{S}^{(1)}_1\mathcal{S}^{(0)}_1\right)\rho_{I(0)}+\mathcal{O}(g_c^8/\gamma_{min}^7).\notag
\end{align}
The terms with a single $\mathcal{S}^{(0)}_{l}$ (first row of eq. (\ref{eTIHg6})) can be identified as the Markov approximation, whereas terms which contain only $\mathcal{S}^{(k)}_{1}$ (one contraction) form the Born approximation. By comparing the order of magnitude of $\mathcal{S}^{(0)}_2=\mathcal{O}(g_c^4/\gamma_{min}^{3})$ and $\mathcal{S}^{(1)}_1\mathcal{S}^{(0)}_1=\mathcal{O}(g_c^4/\gamma_{min}^{3})$, it is clear, that the second order term in the Born expansion is exactly of the same order of magnitude then one of the second order terms in the Markov expansion. This is also valid for all higher order terms. A higher order term in Born always corresponds to a higher cross term in Markov. In this manner a non-Markovian calculation with Born approximation is not reasonable.

A great advantage of our method is that the different terms can be understood by there origin and are anyhow simple to derive. To illustrate how our method can be used to calculate the dynamics of a system we investigate the behavior the spin boson model. A very useful property of our method is that for numerical simulations the terms $\mathcal{S}^{(k)}_{l}$ can be pre-calculated and do not have to be computed every time step.

\section{Spin-boson model}
\label{cSpin}
The 'Drosophila' for the open quantum systems community to check new ideas and expansions is the spin boson model \cite{1-s2.0-S0301010403005469-main,1211.1201}. It explains many interesting problems like electron transfer reactions \cite{Fs10910-012-0124-5}, bio molecules \cite{1.449017}, cavity-QED \cite{PhysRevB.65.235311,PhysRevB.83.180505} and general dissipative quantum systems \cite{PhysRevB.71.035318,1307.3191}. This makes it the perfect choice for checking new models or approximations. The possibility to solve it exact within the Born-approximation \cite{PhysRevB.79.125317} and to solve it perturbatively in a wide parameter regime \cite{0707.0210} further increases its popularity.

The system Hamiltonian is given by $H_S(t)=\tfrac{1}{2}\Delta E\sigma_z+g_D\sigma_x\cos(\omega_D t)\cdot f(t)$. The driving frequency $\omega_D$  is fixed to $\omega_D=\Delta E$, the energy splitting of the qubit and $\sigma_z$ and $\sigma_x$ are Pauli matrices. The function $f(t)$ characterizes the shape of the driving pulse. The bath of harmonic oscillators is described by 
\begin{align}
  H_B=\sum_{i}\omega_i b_i^{\dagger} b_i
\end{align}
with bosonic creation $b_i^{\dagger}$ and annihilation $b_i$ operators. We use a specific coupling to the bath $H_C$ given by
\begin{align}
  H_C=g_c\cdot\sum_{i}\left(\sigma_+ b_i+\sigma_-b_i^{\dagger}\right).
\end{align}
For the treatment of the time dependent part of the system Hamiltonian, it can be useful to change to the rotating frame. This is done by separating the driving from the time independent Hamiltonian, i.e. $\tilde{A}(t)=e^{-\tfrac{i}{2}\Delta E\sigma_zt}A(t)e^{\tfrac{i}{2}\Delta E\sigma_zt}$. 

With an external driving we change also the derivatives of $\tilde{\rho}$ , thus we use eq.(8) of our Markov expansion which still includes these derivatives
\begin{align}
  \tilde{\rho}_{(1)}=i[g_D\sigma_xf(t),\tilde{\rho}_{(0)}]+\sum_{l=0}^{\infty}\tilde{\mathcal{S}}^{(l)}\tilde{\rho}_{(l)}.
  \label{eME}
\end{align}
The derivatives of the reduced density matrix lead to the qubits inertia when it reacts on the external driving. We further assume that the $k^{\text{th}}$ derivative $\tilde{\rho}_{(k)}$ is of order $\mathcal{O}(g_c^{k\cdot2}/\gamma_{min}^k)$, which is exact for a time independent system Hamiltonian (see eq.(21)). Thus, it is important that the driving does not change the system too rapidly, so that our expansion parameter $\mathcal{O}\left(g_c^{2l}/\gamma_{min}^{2l-1+k}\right)$ is still valid.
Then, the QME up to order $\mathcal{O}(g_c^6/\gamma_{min}^5)$ is
\begin{align}
  \tilde{\rho}_{(1)}=&i[\tilde{H}_D(t),\tilde{\rho}_{(0)}]+\left(\tilde{\mathcal{S}}^{(0)}_1+\tilde{\mathcal{S}}^{(0)}_2+\tilde{\mathcal{S}}^{(0)}_3\right)\tilde{\rho}_{(0)}\notag\\
  &+\left(\tilde{\mathcal{S}}^{(1)}_1+\tilde{\mathcal{S}}^{(1)}_2\right)\tilde{\rho}_{(1)}+\tilde{\mathcal{S}}^{(2)}_1\tilde{\rho}_{(2)}+\mathcal{O}(g_c^8/\gamma_{min}^{7})
  \label{eTHg6}
\end{align}
The driving will in principle also change the energy splitting of the qubit. However, we always choose the driving strength $g_D$ to be smaller then the energy splitting of the qubit, $g_D\ll\Delta E$. The energy eigenvalues of the system with driving are $\pm\sqrt{\Delta E^2/4+g_D^2}\approx\pm\tfrac{1}{2}|\Delta E|$. The driving strength we will consider in the simulation is $g_D=0.2\Delta E$ which leads to an energy splitting $\approx\pm0.54|\Delta E|$. Therefore, neglecting the effect of the driving on the energy splitting will not change the expansion of the kernel. 
For this system a contraction in the self-energy $\tilde{\Sigma}$ always contains one raising $\tilde{\sigma}^+$ and one lowering operator $\tilde{\sigma}^-$. An exemplary diagram with one contraction in this picture is given by
\begin{align}
    \begin{tikzpicture}[anchor=base,baseline=8pt]
    \coordinate[] (A) at (0,0);
    \coordinate[label=right:$q'$] (B) at (1,0.1);
    \coordinate[label=left:$\bar{q}$] (C) at (0,0.78);
    \coordinate[] (D) at (1,0.7);
    \node[draw,circle, inner sep=0pt, minimum size=7pt,label=left:$\bar{q}'$,label=below:$t'$] (t1) at (A) {$+$};
    \node[draw,circle, inner sep=0pt, minimum size=7pt,label=right:$q$,label=above:$t$] (t2) at (D) {$-$};
    \node[draw,circle, inner sep=0pt, minimum size=7pt,label=left:$\bar{q}'$,label=below:$t'$] (t1) at (A) {$+$};
    \node[draw,circle, inner sep=0pt, minimum size=7pt,label=right:$q$,label=above:$t$] (t2) at (D) {$-$};
    \draw[line width=1.0pt] (t1) -- (B);
    \draw[line width=1.0pt] (t1) -- (t2);
    \draw[line width=1.0pt] (t2) -- (C);
    \end{tikzpicture}
    =&\bra{\bar{q}}\sigma_-e^{-i\Delta Et}\ket{q}\\
    &\bra{q'}\sigma_+e^{i\Delta Et'}\ket{\bar{q}'}g_c^2\sum_i\langle \tilde{b}_i^{\dagger}(t)\tilde{b}_i(t')\rangle_B.\notag
\end{align}
As a remark, for our one qubit system it is clear that in this diagram $\bar{q}$ has to be the up state $\ket{\uparrow}$, $q$ the down state $\ket{\downarrow}$,  $q'$ the down state $\ket{\downarrow}$ and  $\bar{q}$ the up state $\ket{\uparrow}$.

We define the correlation function according to reference \cite{bStatMethQO} as
\begin{align}
  \sum_i\langle \tilde{b}_i^{\dagger}(t)\tilde{b}_i(t')\rangle_B&=\,C^-(t'-t)\notag\\
    &=\int_0^{\infty}\text{d}\omega J(\omega)n^-(\omega)e^{i\omega(t'-t)}\notag\\
  \sum_i\langle \tilde{b}_i(t)\tilde{b}_i^{\dagger}(t')\rangle_B&=\,C^+(t'-t)\notag\\
    &=\int_0^{\infty}\text{d}\omega J(\omega)n^+(\omega)e^{-i\omega(t'-t)},
\end{align}
with the spectral density function $J(\omega)$ and the Bose-Einstein statistic $n^-(\omega)=\tfrac{1}{\text{exp}(\hbar\omega/k_BT)-1}$ and $n^+(\omega)=n^-(\omega)+1$. Naturally, this leads to the spectral functions
\begin{align}
  \tilde{C}^{\pm}(\omega)=\,J(\omega)n^{\pm}(\omega)
\end{align}
The spectral density function we use is the Ohmic spectral density with Lorentz-Drude cutoff $J(\omega)=\omega/(1+(\tfrac{\omega}{\omega_C})^2)$ with $\omega_C$ the cutoff frequency. For all numerical simulations we set the qubit energy splitting to one, so that all other energies are measured in multiple of $\Delta E$. The inverse temperature is than always $\beta=\tfrac{1}{k_BT}=10\,\Delta E$, the cutoff frequency is $\omega_C=10\,\Delta E$, so that influence is small.

For our specific choice of the system and approximations the calculation of the diagrams to any specific order can be achieved. Because we investigate only one qubit one energy splitting $\Delta E$ between different system states is possible. As described in the main part, the number of integrals in the QME is given by the diagram with $n_c$ contractions containing $2n_c-2$ time ordered integrals caused by the number of inner vertices, one integral from integration of the kernel itself and $k$ integrals from the number of integration by parts (the Markovian order of the term). So, for each diagram $2\,n_c-1+k$ integrals have to be solved. One contraction yields the factors
\begin{align}
  t_l<t_k:\quad &e^{ib\Delta E(t_l-t_k)}C^{c}(d(t_l-t_k)),\notag\\
  &\{b,c,d\}\in\{-1,+1\}
  \label{eCont}
\end{align}
The general form of the integral ordered by the involved times using the limit $t_0\rightarrow-\infty$ of an arbitrary diagram with $n_c$ contractions is given by
\begin{align}
  &\prod_{j=1}^{2n_c-1}\int_{\text{d}\omega_j}I_{n_c}^j=\prod_{j=1}^{2n_c-1}\int_{\text{d}\omega_j}\int_{-\infty}^{t_{j+1}}\text{d}t_j \left(\int_{-\infty}^{t_{2n_c}-t_1}\right)^k\notag\\
  &\,\cdot e^{i a_j t_j(b_j \Delta E-c_jd_j\omega_j)}\cdot e^{ia_{2n_c}t_{2n_c}(b_{2n_c} \Delta E-c_{2n_c}d_{2n_c}\omega_{2n_c})},
  \label{eDiaIntegral}
\end{align}
where the different parameters and constraints on them are given below. The parameter $a_j$ is the sign of $t_j$ from equation (\ref{eCont}). Each vertex from left to right is a time $t_j$, $j=1,2,...$ assigned with $t_j<t_{j+1}$. The frequency integrals arise from the Fourier-transformation of the correlation functions $C^{c_j}(d_j(t_k-t_l))$
\begin{align}
  \int_{\text{d}\omega_j}=\int_{-\infty}^{\infty}\text{d}\omega_jJ(\omega_j)n^{c_j}(\omega_j).
\end{align}
Therefore, it is clear that only $n_c$ Fourier-transformed contractions exist and thus only $n_c$ different $\omega_j$. The integrals will naturally be evaluated only once and not be double counted. The time integrals to the power $k$ are symbolic for the antiderivatives of the kernel.

For a given contraction between the time steps $t_j$ and $t_l$ holds the constraints:
\begin{align}
  \omega_j=\omega_l,\,a_j=-a_l,\,b_j=b_l,\,c_j=c_l,\,d_j=d_l.
\end{align}
By introducing the function
\begin{align}
    \Gamma_j=a_j(b_j\Delta E-c_jd_j\omega_j-ia_j\eta),
\end{align}
the integrals in (\ref{eDiaIntegral}) can be evaluated to ($0<\eta\ll1$ as convergence factor for $t_0\rightarrow-\infty$)
\begin{align}
     \prod_{j=1}^{2n_c-1}I_{n_c}^j=\frac{\text{exp}\left[i\sum_{l=1}^{2n_c}\Gamma_l\,t_{2n_c}\right]}{\prod_{j=1}^{2n_c-2}(\sum_{l=1}^{j}i\Gamma_l)\cdot (\sum_{l=1}^{2n_c-1}i\Gamma_l)^{k+1}}
\end{align}
For each contraction between $t_j$ and $t_l$ the corresponding $\Gamma_j$ and $\Gamma_l$ fulfill the condition $\Gamma_j=-\Gamma_l$ in the limit $\eta\rightarrow0$. Therefore, the exponent vanishes in this limit and the numerator is $1$. The denominator results with the Sokhotsky-Weierstrass theorem
\begin{align}
  \lim_{\eta\rightarrow0^+}\frac{1}{(x+i\eta)^n}=\text{P}\frac{1}{x^n}-i\pi\tfrac{(-1)^{n-1}}{(n-1)!}\delta^{(n-1)}(x),
\end{align}
where P denotes a principal value integral, in the solution
\begin{align}
  \prod_{j=1}^{2n_c-1}&I_{n_c}^j=\prod_{j=1}^{2n_c-2}(-1)^{n_c+1}[\pi\delta^{0}(\sum_{l=1}^{j}-\text{Re}\{\Gamma_l\})\notag\\
    &+i\text{P}\frac{1}{\sum_{l=1}^{j}-\text{Re}\{\Gamma_l\}}]\notag\\
    &\cdot\left((-1)^{n_c+1}[\pi\tfrac{(-1)^{k}}{k!}\delta^{(k)}(\sum_{l=1}^{2n_c-1}-\text{Re}\{\Gamma_l\})\right.\notag\\
    &\left.+i\text{P}\frac{1}{(\sum_{l=1}^{2n_c-1}-\text{Re}\{\Gamma_l\})^{k+1}}]\right).
  \label{eSelfEv}
\end{align}
The next step is to add up diagrams that contribute to the same in and outgoing states and the same correlations in between. For example, inversion in the center of the diagrams always leads to such behaviour. This corresponds to changing the sign of $a_j$, but not touching the other parameters $b_j$, $c_j$ and $d_j$. This inversion yields a changing sign of the principal value part in equation (\ref{eSelfEv}) and therefore vanishing imaginary part of the self energy.

For one contraction only the real part with one delta-distribution $\delta(\omega-\Delta E)$ remains. For more contractions the real part contains always one term with only delta distributions and no principal value which leads to the evaluation of the spectral functions at the qubit frequency $\Delta E$, but in principal combinations of the principal value integrals also result in real terms. By adding all diagrams belonging to one element of the reduced density matrix, these terms cancel, because the parameter $c_j$ is the same for all this diagrams and the parameters $a_j$, $b_j$ and $d_j$ lead to all combination of sign changes that are than added up. The example below for two contractions shows the idea. 
{\allowdisplaybreaks
\begin{align}
 &\begin{tikzpicture}[anchor=base,baseline=8pt]
    \coordinate (A) at (0,0);
    \coordinate (B) at (1,0);
    \coordinate[label=above:{\tiny $+$}] (C) at (0,0.7);
    \coordinate[label=above:{\tiny $-$}] (D) at (1,0.7);
    \coordinate[label=below:{\tiny $+$}] (E) at (0.6,0.0);
    \coordinate[label=below:{\tiny $-$}] (F) at (0.3,0.0);
    \draw[line width=1pt] (A) -- (B);
    \draw[line width=1pt] (C) to [bend right=70](D);
    \draw[line width=1pt] (F) to [bend left=80](E);
    \draw[line width=1pt] (D) -- (C);
    \fill (C) circle (2pt);
    \fill (D) circle (2pt);
    \fill (E) circle (2pt);
    \fill (F) circle (2pt);
  \end{tikzpicture}+
   \begin{tikzpicture}[anchor=base,baseline=8pt]
    \coordinate[label=below:{\tiny $-$}] (A) at (0,0);
    \coordinate[label=below:{\tiny $+$}] (B) at (1,0);
    \coordinate (C) at (0,0.7);
    \coordinate (D) at (1,0.7);
    \coordinate[label=above:{\tiny $-$}] (E) at (0.6,0.7);
    \coordinate[label=above:{\tiny $+$}] (F) at (0.3,0.7);
    \draw[line width=1pt] (A) -- (B);
    \draw[line width=1pt] (B) to [bend right=70](A);
    \draw[line width=1pt] (F) to [bend right=80](E);
    \draw[line width=1pt] (D) -- (C);
    \fill (A) circle (2pt);
    \fill (B) circle (2pt);
    \fill (E) circle (2pt);
    \fill (F) circle (2pt);
  \end{tikzpicture}+
     \begin{tikzpicture}[anchor=base,baseline=8pt]
    \coordinate (A) at (0,0);
    \coordinate[label=below:{\tiny $+$}] (B) at (1,0);
    \coordinate[label=above:{\tiny $+$}] (C) at (0,0.7);
    \coordinate (D) at (1,0.7);
    \coordinate[label=above:{\tiny $-$}] (E) at (0.6,0.7);
    \coordinate[label=below:{\tiny $-$}] (F) at (0.3,0.0);
    \draw[line width=1pt] (A) -- (B);
    \draw[line width=1pt] (C) to [bend right=70](E);
    \draw[line width=1pt] (B) to [bend right=80](F);
    \draw[line width=1pt] (D) -- (C);
    \fill (B) circle (2pt);
    \fill (C) circle (2pt);
    \fill (E) circle (2pt);
    \fill (F) circle (2pt);
  \end{tikzpicture}+
     \begin{tikzpicture}[anchor=base,baseline=8pt]
    \coordinate[label=below:{\tiny $-$}] (A) at (0,0);
    \coordinate (B) at (1,0);
    \coordinate (C) at (0,0.7);
    \coordinate[label=above:{\tiny $-$}] (D) at (1,0.7);
    \coordinate[label=below:{\tiny $+$}] (E) at (0.6,0.0);
    \coordinate[label=above:{\tiny $+$}] (F) at (0.3,0.7);
    \draw[line width=1pt] (A) -- (B);
    \draw[line width=1pt] (A) to [bend left=70](E);
    \draw[line width=1pt] (F) to [bend right=80](D);
    \draw[line width=1pt] (D) -- (C);
    \fill (A) circle (2pt);
    \fill (D) circle (2pt);
    \fill (E) circle (2pt);
    \fill (F) circle (2pt);
  \end{tikzpicture}=\notag\\
  &\int_{\omega_1}\int_{\omega_2}\left[\left(\pi\delta(\omega_1-\Delta E)+\text{P}\tfrac{i}{\omega_1-\Delta E}\right)\right.\notag\\
  &\left(\pi\delta(\Delta E-\omega_2)+\text{P}\tfrac{i}{\Delta E-\omega_2}\right)\left(\pi\delta(\omega_1-\omega_2)+\text{P}\tfrac{i}{\omega_1-\omega_2}\right)+\notag\\
  &\left(\pi\delta(\omega_1-\Delta E)-\text{P}\tfrac{i}{\omega_1-\Delta E}\right)\left(\pi\delta(\Delta E-\omega_2)-\text{P}\tfrac{i}{\Delta E-\omega_2}\right)\notag\\
  &\left(\pi\delta(\omega_1-\omega_2)-\text{P}\tfrac{i}{\omega_1-\omega_2}\right)+\notag\\
  &\left(\pi\delta(\Delta E-\omega_1)+\text{P}\tfrac{i}{\Delta E-\omega_2}\right)^2\left(\pi\delta(\omega_2-\omega_1)+\text{P}\tfrac{i}{\omega_2-\omega_1}\right)+\notag\\
  &\left.\left(\pi\delta(\Delta E-\omega_1)-\text{P}\tfrac{i}{\Delta E-\omega_2}\right)^2\left(\pi\delta(\omega_2-\omega_1)-\text{P}\tfrac{i}{\omega_2-\omega_1}\right)\right]\notag\\
  &=\int_{\omega_1}\int_{\omega_2}4\pi^3\delta(\cdots)^3\notag\\
  &+\int_{\omega_1}\int_{\omega_2}2\pi\left(\underbrace{-\tfrac{\text{P}\delta(\omega_1-\omega_2)}{(\omega_1-\Delta E)^2}+\tfrac{\text{P}\delta(\omega_1-\omega_2)}{(\omega_1-\Delta E)(\omega_2-\Delta E)}}_{=0}\right.\notag\\
  &\left.-\underbrace{\tfrac{\text{P}\,2\,\delta(\omega_1-\Delta E)}{(\omega_1-\Delta E)(\omega_1-\omega_2)}}_{=0(\text{PV})}\underbrace{-\tfrac{\text{P}\delta(\Delta E-\omega_2)}{(\omega_1-\Delta E)(\omega_1-\omega_2)}+\tfrac{\text{P}\delta(\omega_1-\Delta E)}{(\omega_2-\Delta E)(\omega_1-\omega_2)}}_{=0}\right)
\end{align}
}
The rules to evaluate a diagram are inspired by reference \cite{PhysRevB.50.18436}, but now specific for our system:
\begin{enumerate}
  \item A contraction from a \textcircled{$\mathsmaller{-}$}-Vertex to a \textcircled{$\mathsmaller{+}$}-Vertex along the Keldysh-contour gives a factor $\tfrac{\partial^k\tilde{C}^-(\pm\Delta E)}{\partial \omega^k}$
  \item A contraction from a \textcircled{$\mathsmaller{+}$}-Vertex to a \textcircled{$\mathsmaller{-}$}-Vertex along the Keldysh-contour gives a factor $\tfrac{\partial^k\tilde{C}^+(\pm\Delta E)}{\partial \omega^k}$
  \item The prefactor $g_c^{2n_c}\cdot(-1)^{n_c+b}$ is given by $n_c$ the number of contractions, $b$ the number of vertices on the lower contour and $g_c$ the coupling constant to the bath.
  \item Each vertex gives a factor $\bra{\bar{q}}\sigma_i\ket{q}$ with $\bar{q}$ the incoming state and $q$ the outgoing state.
\end{enumerate}

We investigate the different decay of an excited state of a non-Born-Markov (NBM) simulation including initial correlations with a simulation using the Born-Markov (BM) approximation and only using the Born approximation. The QME in the BM approximation is given by
\begin{align}
  \tilde{\rho}_{(1)}=\tilde{\mathcal{S}}^{(0)}_1\tilde{\rho}_{(0)},
\end{align}
which is equivalent to \cite{bStatMethQO}
\begin{align}
 \dot{\tilde{\rho}}(t)=-\int_0^{t}\text{d}\tau\, \text{tr}_B\left\{\left[H_C(t),[H_C(\tau),\tilde{\rho}(t)\rho_B]\right]\right\}.
\end{align}
The Born approximation to all orders without Markov is given by eq. (\ref{eExpME}) with only single contraction diagrams, i.e. $\mathcal{S}^{(l)}_1$ terms
\begin{align}
  \tilde{\rho}_{(1)}(t)=\sum_{l=0}^{\infty}\mathcal{S}^{(l)}_1\tilde{\rho}_{(l)}(t),
\end{align}
which is in all orders equivalent to reference \cite{bStatMethQO}
\begin{align}
 \dot{\tilde{\rho}}(t)=-\int_0^{t}\text{d}\tau\, \text{tr}_B\left\{\left[H_C(t),[H_C(\tau),\tilde{\rho}(\tau)\rho_B]\right]\right\}.
\end{align}
We compare the dynamics up to the third order in the Markovian expansion
\begin{align}
 \tilde{\rho}_{(1)}=\tilde{\mathcal{S}}^{(0)}_1\tilde{\rho}_{(0)}+\tilde{\mathcal{S}}^{(1)}_1\tilde{\rho}_{(1)}+\tilde{\mathcal{S}}^{(2)}_1\tilde{\rho}_{(2)}.
 \label{eBorn}
\end{align}
To include initial state correlations of the excited state we start our simulation with an equilibrium state and use a weak $\pi/2$-pulse in the rotating frame to excite the system. In this setup the initial correlations appear naturally caused by the preparation. The shape of the $\pi/2$-pulse is given by
\begin{align}
  f(t)=\Theta(t-t_p-\tfrac{\pi}{2g_D})\Theta(t_p-t).
\end{align}
The parameter $t_p$ is the end of the pulse, the length of the pulse is $\tfrac{\pi}{2g_D}$, the height is the driving strength $g_D$, so that the area under the pulse is exactly $\pi/2$. 

We pulse the NBM system to an excited state which we then also use as the starting point for the BM simulation and Born simulation. Thus, we start our investigation of the decay in all simulations with the same state in which the initial state correlations are included.
To have a mechanism to measure the difference of the two density matrices, the trace distance has the right properties to do so \cite{0908.0238}. It is defined as 
\begin{align}
  D(A,B)=\tfrac{1}{2}||A-B||_1,
\end{align}
where $A$ and $B$ are two trace class operators and $||\,\,||_1$ is the trace norm. For our purpose it is important to get information about the distinguishability between two reduced density matrices which is exactly the physical interpretation of the trace distance \cite{0908.0238}. Furthermore, it can be used as a measure for the strength of the non-Markovian behaviour \cite{0908.0238} by testing the increase in time of the trace distance for two reduced density matrices. This corresponds to a back flow of information from the bath.

\begin{figure}[htp]
  \includegraphics[origin=c,width=0.48\textwidth]{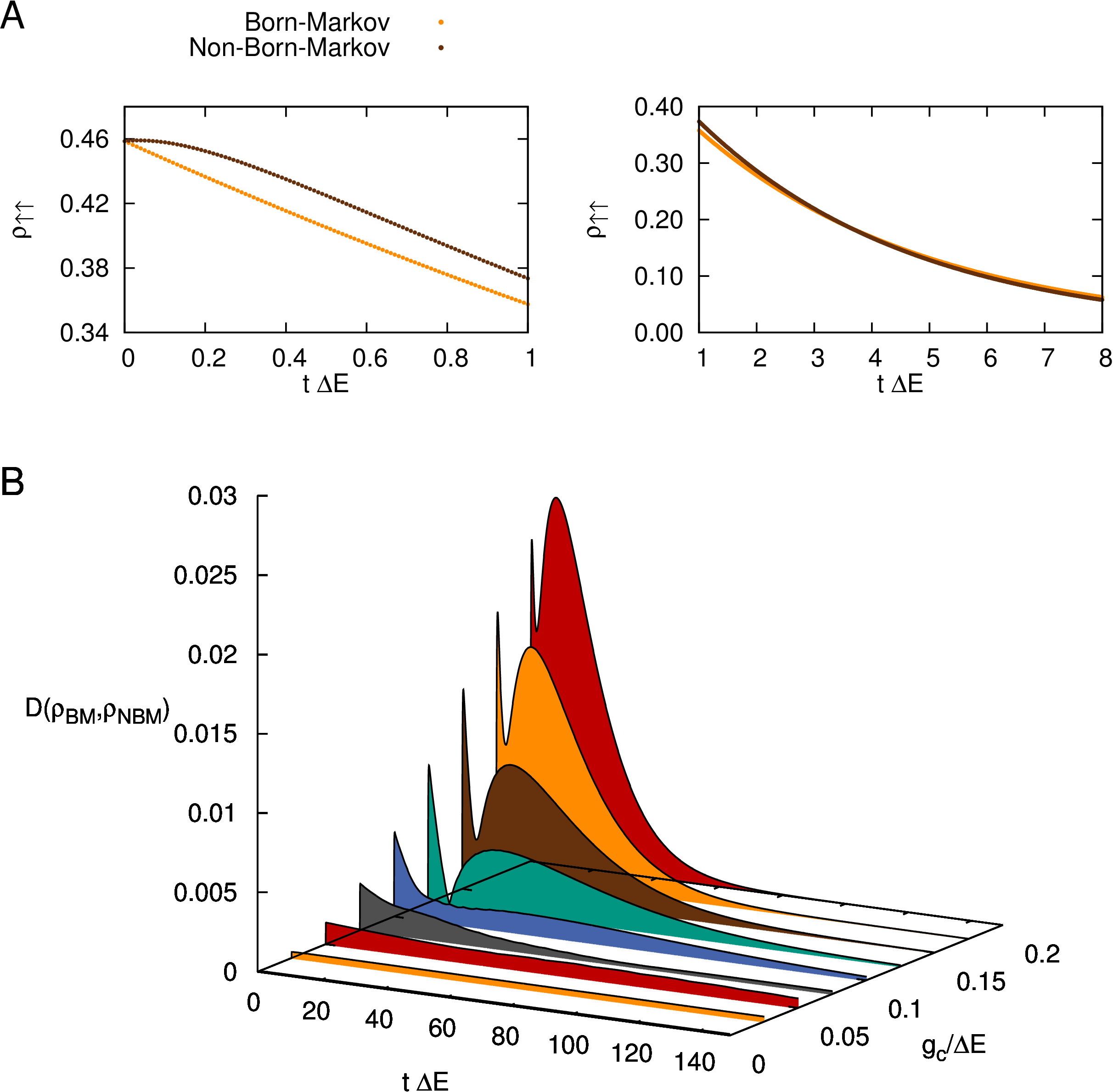}
  \caption{In all figures $t=0$ corresponds to the end of the pulse $t_p$. (A) Comparison of the decay of the excited state between simulations with and without BM-approximation for the strongest coupling $g_c=0.2$ for short times and intermediate times. (B) The trace distance $D(\rho_{BM},\rho_{NBM})$ shows the distinguishability between reduced density matrices of the BM simulation $\rho_{BM}$ and the NBM simulation $\rho_{NBM}$ for different coupling strengths $g_c$. The system relaxes in its equilibrium steady state which is approximately the ground state of the system for $\beta=10\,\Delta E$. The pulse strength is set to $g_d=0.2\,\Delta E$.}
  \label{pDecay}
\end{figure}
First, we compare our expansion with the Born-Markov approximation. The figure \ref{pDecay} (A) shows the immediate exponential decay for the BM simulation, where on the other side the NBM simulation is depending on its initial correlations and thus on its past. For short to intermediate times this leads to different dynamics. In figure \ref{pDecay} (C) the trace distance starts for all couplings at zero but rises rapidly to its peak for very short times. The distinguishability then decays for all coupling strength, but for stronger couplings a local minimum is reached for intermediate times. This can be interpreted as a back flow of information in the NBM case and therefore is a measure for the non-Markovianity of the system. This back flow is the larger the stronger the coupling is to the bath. For long times the system decays in its ground state and than gets indistinguishable.
So, for the decay of the one qubit system, the higher order terms get more important for stronger couplings, as known, and the non-Markovian back flow of information can be seen.

\begin{figure}[htp]
  \includegraphics[origin=c,width=0.48\textwidth]{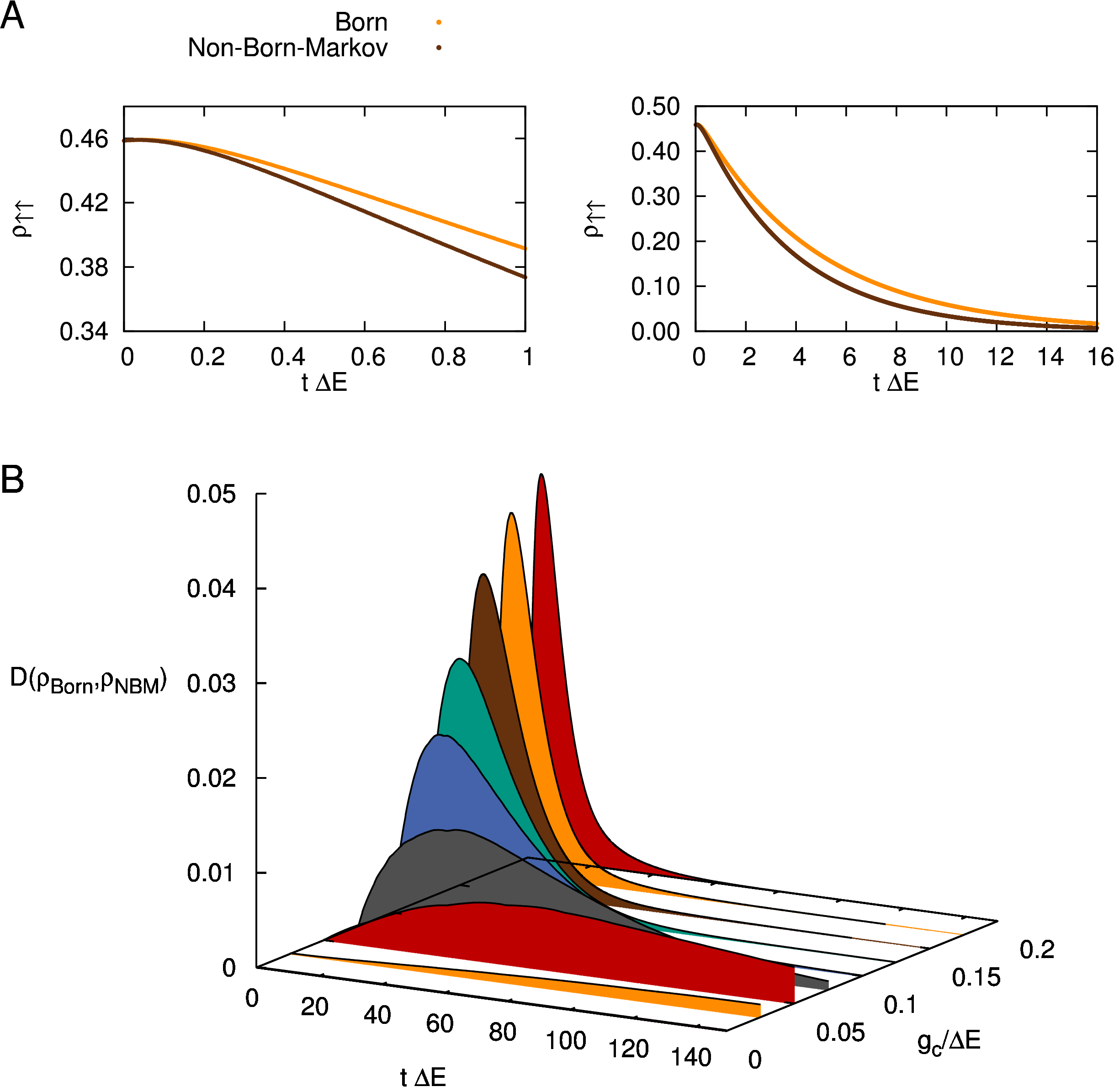}
  \caption{In all figures $t=0$ corresponds to the end of the pulse $t_p$. (A) Comparison of the decay of the excited state between simulations with and without Born approximation for the strongest coupling $g_c=0.2$ for short times and intermediate times. (B) The trace distance $D(\rho_{Born},\rho_{NBM})$ shows the distinguishability between reduced density matrices of the Born simulation $\rho_{Born}$ and the NBM simulation $\rho_{NBM}$ for different coupling strengths $g_c$. The system relaxes in its equilibrium steady state which is approximately the ground state of the system for $\beta=10\,\Delta E$. The pulse strength is set to $g_d=0.2\,\Delta E$.}
  \label{pBorn}
\end{figure}
Second, we investigate the importance of the higher order Born terms (see figure \ref{pBorn}) in our expansion. Without the higher order terms (see eq. (\ref{eBorn})), the decay of the system is too slow, as shown in the figures \ref{pBorn} (A). As one would  suspect, the stronger the coupling is the bigger is the influence of higher order terms and the trace distance gets larger. Surprisingly, the difference between a simulation with Born approximation, but without Markov approximation and the full expansion is even larger for the strong coupling as the difference to the Born-Markov approximation. The higher order Markov terms slow the decay of the excited state down, whereas higher order Born terms lead to a faster decay. This supports the importance of adding up all terms belonging to the same order of magnitude.

We want to emphasize here again, that the power of our method is not to calculate the dynamics of the QME explicit like it was done in this part, but to analyze open problems by their order of magnitude. The calculation of higher order terms in the Born expansion, so higher order diagrams is in general challenging. But the advantage of our method in comparison to directly calculating higher order diagrams of the QME in its original form is that the final terms are not time dependent anymore and can be efficiently pre-calculated for numerical simulations and do not have to be changed every time step. 

\section{Initial state problem}
\label{cIni}
To show that our method is useful for other problems without calculating kernels explicit, we investigate the initial state problem occurring in non-Markovian dynamics. Instead of using the limit $t_0\rightarrow -\infty$, it is common to start with an initial non-equilibrium state at time $t_0=0$ and investigate the resulting dynamics. To get a proper understanding of the behavior of the QME for such an initial condition, we cut the integral of the exact description for $t_0\rightarrow -\infty$ at $t_c=0$ 
\begin{align}
  \dot{\rho}_I=\underbrace{\int_{-\infty}^0 \text{d}t' \Sigma_I^{(0)}(t-t')\rho_I(t')}_{A_0}+\underbrace{\int_{0}^t \text{d}t' \Sigma_I^{(0)}(t-t')\rho_I(t')}_{B_0}.
\end{align}
The term $A_0$ represents the initial correlations of the system. The initial correlations have an effect on the dynamics of the system which becomes stronger as the system becomes more non-Markovian. Therefore, the next step of our analysis is to apply the Markov expansion as described before to both terms. The $k^{\text{th}}$ order of the Markov expansion meaning $k$ integrations by part of $A_0$ and $B_0$ are
\begin{widetext}
\begin{align}
  &A_k=\int_{-\infty}^0 \text{d}t'\Sigma_I^{(k)}(t-t')\rho_{I(k)}(t')=\underbrace{\left(\int_{\infty}^t \text{d}t'\Sigma_I^{(k)}(t')\right)\rho_{I(k)}(0)}_{A_k^{IC}}+\underbrace{\int_{-\infty}^0\text{d}t'\Sigma_I^{(k+1)}(t-t')\rho_{I(k+1)}(t')}_{A_{k+1}}\\
  &B_k=\int_{0}^t \text{d}t' \Sigma_I^{(k)}(t-t')\rho_{I(k)}(t')=\underbrace{\vphantom{\left(\int_{\infty}^t \text{d}t'\Sigma_I^{(k)}(t')\right)}\mathcal{S}^{(k)}\rho_{I(k)}(t)}_{\text{see eq. (\ref{eExpME})}}-\underbrace{\left(\int_{\infty}^t \text{d}t'\Sigma_I^{(k)}(t')\right)\rho_{I(k)}(0)}_{B_k^{IC}}+\underbrace{\int_{0}^t\text{d}t'\Sigma_I^{(k+1)}(t-t')\rho_{I(k+1)}(t')}_{B_{k+1}}.
   \label{eIniME}
\end{align}
\end{widetext}
A complete non-Markovian simulation from time $t_c=0$ corresponds to considering the terms $B_k$ to all orders and neglecting all $A_k$. This means, that in each order $k$ the correct term $\mathcal{S}^{(k)}\rho_{I(k)}(t)$ is added, but also an unwanted term $B_k^{IC}$ produced by the initial correlations. These initial correlations are decaying exponentially like the kernel $\Sigma_I^{(k)}(t)$ with the correlation time of the bath. Thus, the $B_k^{IC}$ for times $t$ larger then $1/\gamma_{min}$ go to zero, but can be important for the short time behavior.

The terms $A_k^{IC}$ and $B_k^{IC}$ are identical. Summing up the terms $A_k$ and $B_k$ to all orders yields the exact limit $t_0\rightarrow -\infty$ without cut, since the terms $A_k^{IC}$ and $B_k^{IC}$ cancel each other in all orders. With our method it is possible to calculate the effect of initial correlations to any order.

\section{Conclusion}
\label{cCon}
We developed another and less complex way to find the exact expansion of the QME in the coupling to the bath and in the bath correlation time for an open quantum system resulting in a time local equation. With this method it is possible to calculate higher order terms in the Born and Markov expansion and distinguish between these. In particular, the order of magnitude of each term can be quantified and shows that higher order terms in the Born expansion are of the same order of magnitude as higher order non-Markovian terms. Therefore, we state that a non-Markovian investigation of a system also requires going beyond Born approximation. Secondly we address the initial state problem of a non-Markovian time evolution. Specific terms can be identified as initial correlations by cutting the exact time evolution and can be calculated to all orders. This result gives scientists an easy tool to estimate the validity of the common Born, Markov and initial state approximation and to go beyond.

{\bf Acknowledgments:} We would like to thank G. Sch\"on, J. Jin, S. Zanker, D. Mendler and A. Heimes for enlightening discussions and support.

\bibliography{/home/chris/bwSyncAndShare/Share/Promotion/Paper/bibliography2}

\end{document}